\documentclass[runningheads]{llncs}
\usepackage{graphicx}
\begin{document}
\title{OpenTable data with multi-criteria ratings}
%
%
\author{Yong Zheng}
\authorrunning{Yong Zheng}
%
\institute{Department of Information Technology \& Management\\
Illinois Institute of Technology, Chicago, IL 60661, USA\\
\email{yzheng66@iit.edu}
}
\maketitle              
\begin{abstract}
With the development of recommender systems (RSs), several promising systems have emerged, such as context-aware RS, multi-criteria RS, and group RS. Multi-criteria recommender systems (MCRSs) are designed to provide personalized recommendations by considering user preferences in multiple attributes or criteria simultaneously. Unlike traditional RSs that typically focus on a single rating, these systems help users make more informed decisions by considering their diverse preferences and needs across various dimensions. In this article, we release the OpenTable data set which was crawled from OpenTable.com. The data set can be considered as a benchmark data set for multi-criteria recommendations.

\end{abstract}
\section{Introduction}

\begin{figure}[h] 
    \centering
    \includegraphics[width=0.7\textwidth]{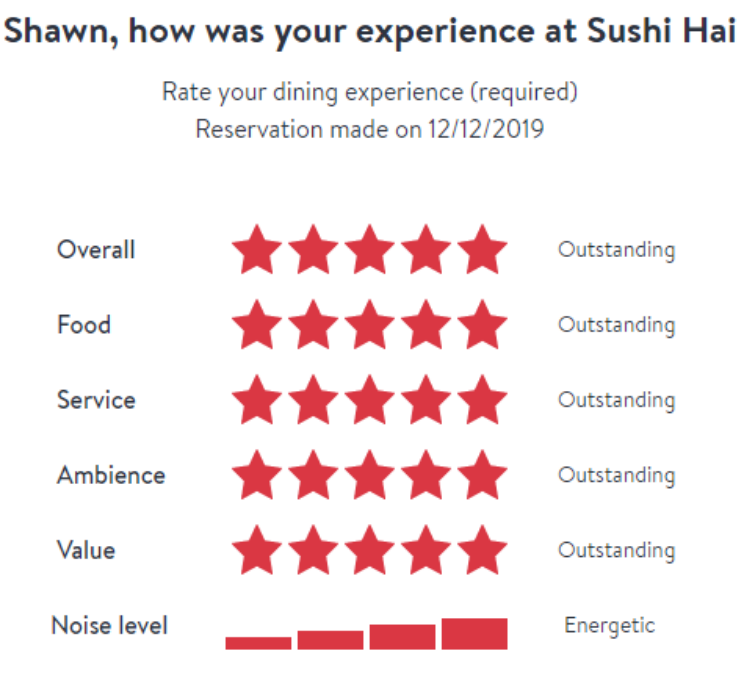} 
    \caption{Example of Ratings on OpenTable.com}
    \label{fig:opentable}
\end{figure}

Recommender systems (RSs) are advanced algorithms that help users discover products or services that align with their preferences by analyzing historical data and user behavior. Multi-criteria recommender systems (MCRSs)~\cite{adomavicius2010multi,zheng2023multi} can further take user preferences in multiple aspects of the items (i.e., criteria) into consideration. For instance, in the context of ratings from OpenTable.com, as shown by Figure~\ref{fig:opentable}, MCRSs can consider user preferencs across various criteria, such as food quality, service, ambience, and value, to generate personalized restaurant recommendations. By considering these multiple factors, MCRSs enhance user satisfaction and decision-making, providing tailored suggestions that reflect the diverse needs of diners.

Accordingly, the rating data on OpenTable.com can be shown by Table~\ref{tab:data}. We may want to predict how user $U_3$ give ratings to the item $T_1$. In this example the column ``rating" refers to a user's overall rating on the item. The values in the last three columns represent a user's ratings on an item from the perspective of multiple criteria, including food quality, satisfaction of services and ambience.

\begin{table}[h] 
    \centering
    \includegraphics[width=0.7\textwidth]{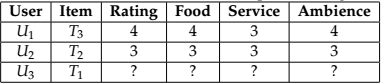} 
    \caption{Example of Multi-Criteria Rating Data} 
    \label{tab:data} 
\end{table}

In the area of MCRSs, there are several well-known data sets, such as the ratings from TripAdvisor.com where we have ratings on hotels from the perspective of room size, room cleanliness, location, check-in experiences and business services, and the rating data from Yahoo!Movies where we have users' ratings on movies in view of visual pictures, quality of story, music effects, and so forth. Though OpenTable.com is another example of real-world applications with multi-criteria ratings, there are no open data sets available for research. 

\section{The OpenTable Data Set}

\subsection{Data Collections}
We utilized Web crawling to acquire restaurant ratings from OpenTable.com, where both the overall rating and multi-criteria ratings are included. The major challenge in the process of data collection is identifying users. There are several anonymous ratings given by users. Namely, we are not able to identify a specific user or UserID from the Webpages. As a result, it is difficult to acquire dense ratings.

\subsection{Data Description and Statistics}
The OpenTable data set has been released on Kaggle.com\footnote{\url{https://www.kaggle.com/datasets/irecsys/opentable-data-with-multi-criteria-ratings}} and IEEE DataPort\footnote{\url{https://ieee-dataport.org/documents/opentable-data-multi-criteria-ratings}}. There are 19,536 ratings given by 1,309 users on 91 restaurants. In addition to the overall ratings, we have users' ratings on the restaurants from 4 criteria, including \textit{food} quality, satisfaction of \textit{service} and \textit{ambience}, and the overall \textit{value} of the picks. The ratings were given in the scale of 1 to 5.

\begin{figure}[h] 
    \centering
    \includegraphics[width=1\textwidth]{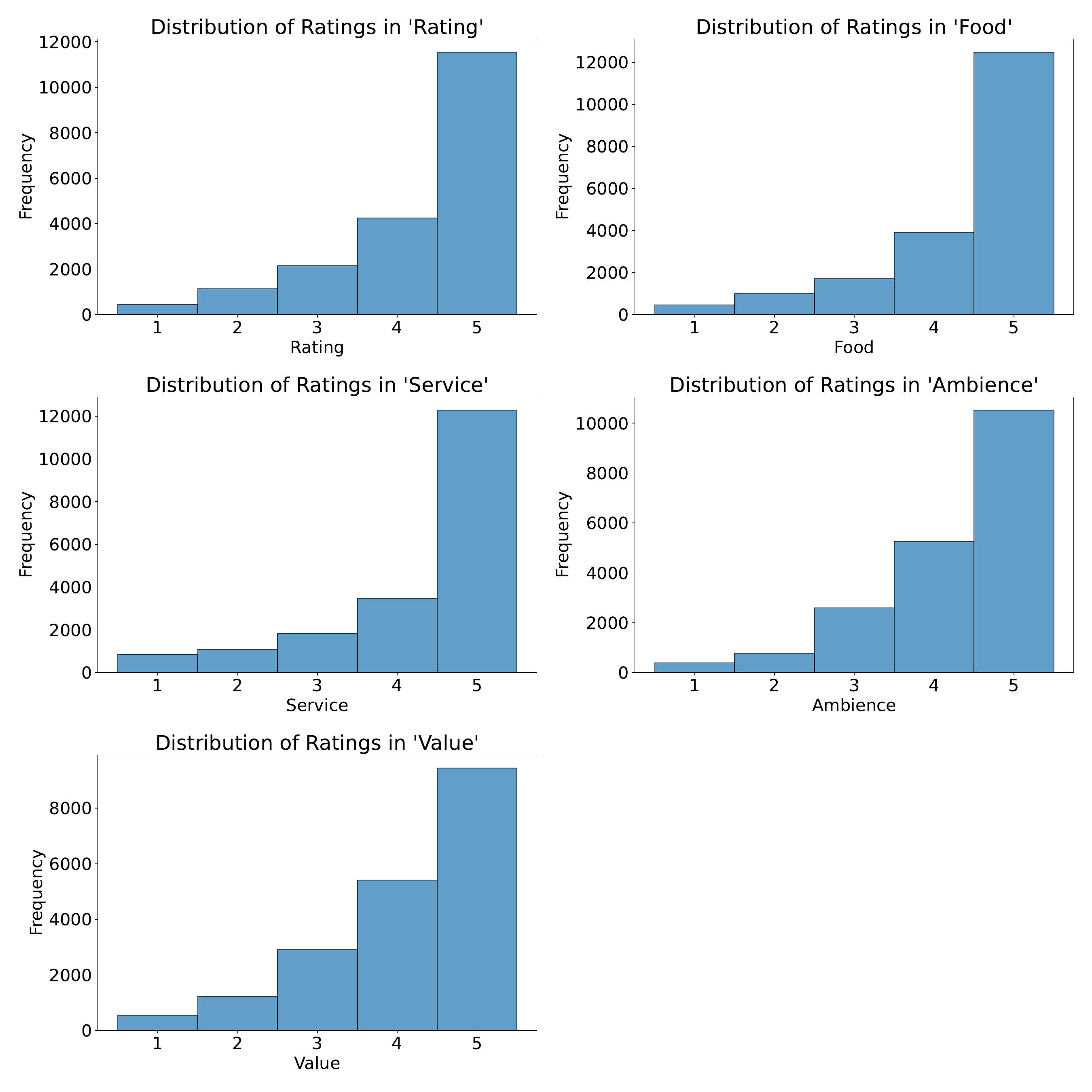} 
    \caption{Rating Distributions}
    \label{fig:dist}
\end{figure}

In terms of the rating distributions on these columns related to user preferences, we present the distribution by using Figure~\ref{fig:dist}.

The rating sparsity is 83.6\%. In terms of the number of ratings per user, the average number is 14.9 and median is 3. Regarding the number of ratings per restaurant, the average and median values are 214 and 250, respectively.

\subsection{Data Usage}
This data set can be used for traditional recommendations by utilizing the overall rating only. It can serve as a benchmark data set for MCRSs by using both the overall rating and the multi-criteria ratings. We have examined different MCRSs algorithms over this data set in our previous work~\cite{zheng2022multi,zheng2023multi,zheng2023hybrid} by using the MCRecKit library~\cite{zheng2024mcreckit}.

\subsection{Special Notes}
While working on web crawling for OpenTable.com, we found that HTML resources contained users' nickname instead of UserIDs. Occasionally, anonymous reviews used a default username, such as "Unknown user." To assign UserIDs, we treated the combination of username and city as a unique user. However, for entries with the default "Unknown user" username, we assigned the same ID. This explains why you might see multiple entries with the same $<$user, item$>$ pair but different ratings. Therefore, we additionally provide the opentable\_cleaned.csv file in the data repository in Kaggle.com and IEEE DataPort, where we removed duplicated entries and only include the last entry associated with the unique $<$user, item$>$ pair in the data set.

%
%
%
\bibliographystyle{splncs04}
\bibliography{sample-base}

\end{document}